# What can Planetary Ice Shells Teach Us about Ocean Mixing and Life, and Why Salt Matters

Nicole Shibley[1,2*]

[1*]Department of Applied Mathematics & Theoretical Physics, University of Cambridge, Wilberforce Road, Cambridge, CB3 0WA, UK. [2]Department of Earth Sciences, University of Cambridge, Downing St, Cambridge, CB2 3EQ, UK.

Corresponding author(s). E-mail(s): ncs32@cam.ac.uk;

**Abstract**

Past and present climates, both on Earth and in planetary systems, maintain ice covers with underlying oceans. This can range from sea ice in the Arctic Ocean to global ice shells on ice-covered moons in the Solar System to ice sheets in past episodes of Earth's history. When an ice shell overlies an ocean, it becomes challenging to infer what ocean processes could be going on below as in-situ ocean measurements are often lacking. This means that what we learn about ocean mixing must often be inferred from above-surface measurements. Here we synthesize the following ideas: (1) surface observations of past and present planetary ice shells give insight into subsurface ocean stratifications and mixing (and vice versa), (2) the stratification, mixing, and presence of ice in (1) is largely controlled by the ocean salinity, and (3) the interplay between (1) and (2) in past and present planetary systems may provide new insights into habitability and the origins of life.

## 1 Ocean Worlds

Several ocean worlds exist in the Solar System [e.g., 1]. These range from our own Earth System to planetary bodies such as Europa and Enceladus. Ice-ocean processes are key to understanding habitability in both Earth and non-Earth planetary environments. On Earth, habitability relates to the conditions, such as surface temperature, which

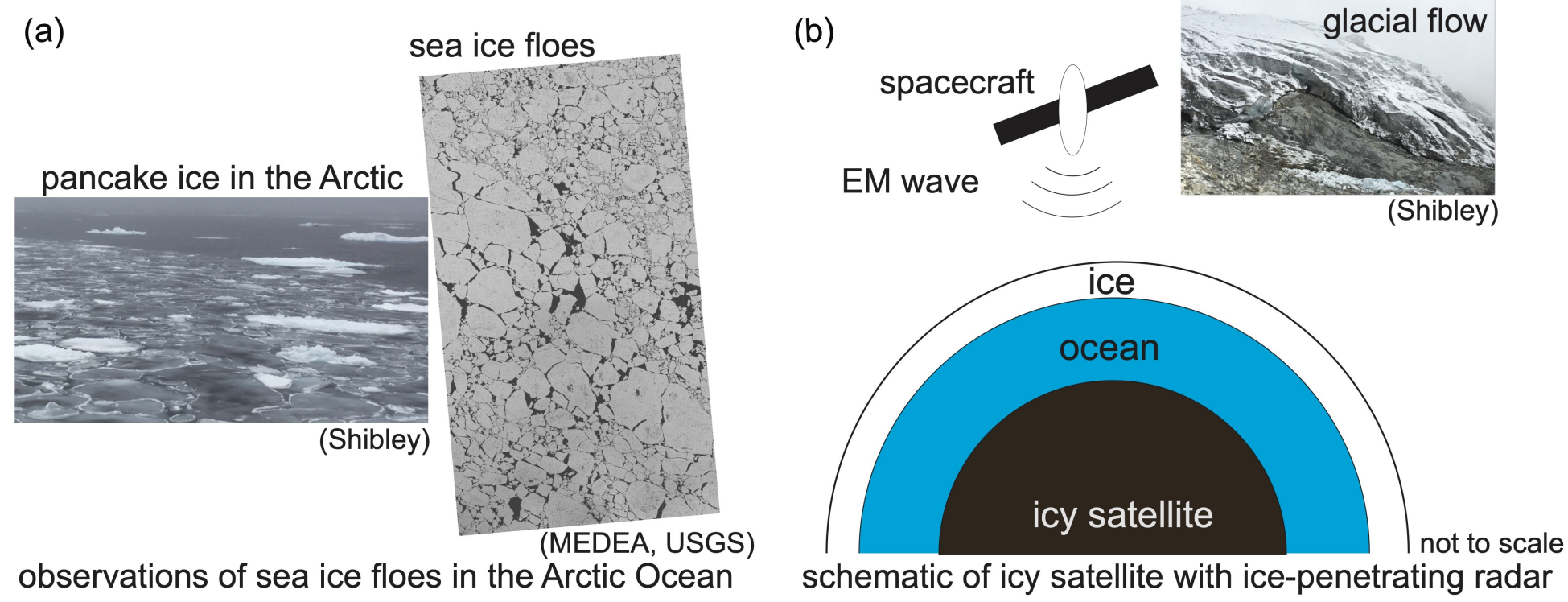

**Fig. 1** (a) Examples of different kinds of ice floes in the Arctic Ocean. The first image is of pancake ice taken by the author during field work. This is early-season ice that is generally pushed together by wind and waves, creating "pancakes." The image on the right is from the USGS MEDEA program, indicating floes of various different sizes. The reflectivity of ice helps regulate Earth's surface temperature. (b) Schematic of an ice-covered Solar System satellite, such as Europa or Enceladus. These satellites are thought to have kilometers-thick ice shells overlying deep oceans. Both Europa *Clipper* and ESA's *Juice* mission will have ice-penetrating radar to image Europan ice thickness. It is thought that such ice shells flow similarly to the movement of glaciers (top right image by the author).

allow for a hospitable planet. Here, polar ice-ocean processes play a pivotal role in regulating Earth's temperature and climate, as ice reflects incoming solar radiation back to space [e.g., 2]; the polar regions are also acutely sensitive to the effects of climate change (Figure 1a). Elsewhere in the Solar System, habitability describes whether a world could sustain life; if it does, where should we look for it? On the Solar System's ice-covered moons thought to have subsurface liquid-water oceans [e.g., 3–5], ice-ocean interactions may shed insight on ocean dynamics and potential locations of possible life [e.g., 6, 7]. How can we learn about ocean mixing by observing ice, and what can ice teach us about ocean mixing in these planetary settings? We synthesize research across past and present climates in the Solar System, blending the idea that in ice-covered systems the salinity stratification is the central to understanding both mixing and habitability.

Perhaps the most explored ice-ocean planetary systems are Europa, one of Jupiter's moons, and Enceladus, one of Saturn's (Figure 1b). Space missions such as *Galileo* and *Cassini*, respectively, have returned various images and measurements from these icy satellites [e.g., 5, 8]. Both bodies are thought to have global, salty oceans sitting below their ice covers [e.g., 1]. How do we know? Evidence for an ocean is supported in part by density inferences drawn from surface gravity measurements, and further buttressed by magnetometer measurements from the *Galileo* mission [4]. This groundbreaking work indicated that magnetic field perturbations could arise from a conductive fluid, like saltwater, subsurface. Further, cracks on surfaces of both systems indicate that there may be an ocean below [5, 9] and imagery from Enceladus displays geysers emanating from the surface [5]. (On Europa, inferences and hypotheses for geysers remain more tenuous [10, 11].) Moreover, observations of orange-colored material on Europa's surface, thought to be due to irradiated salts [12], along with carbon inferences from telescopic data [13], give further credence to the idea that a salty, subsurface ocean lies beneath the ice shell.

## 2 Challenge: Lack of Subsurface Measurements

However, despite significant evidence for planetary oceans, a major challenge in understanding ocean dynamics below ice shells both in planetary and polar Earth settings, is the lack of subsurface measurements [e.g., 14, 15]. On planetary systems, current ocean inferences arise largely from magnetic measurements that are linked to ocean conductivities (and hence salinities) [e.g., 16]. Telescopic measurements demonstrate elemental compositions, such as oxygen or hydrogen [e.g., 17, 18]. And ice-penetrating radar measurements, such as from the upcoming *Clipper* and *Juice* missions [e.g., 19, 20], will allow for inferences of ice thickness (in optimal conditions). While certain missions have been proposed to land on icy satellites and more accurately infer or measure properties of their oceans, such as Europa Lander [21] and Enceladus Orbilander [22], as of yet, none have. Therefore, since the ice and ocean are dynamically and thermodynamically linked [e.g., 15, 23], we can utilize the ice cover as the principal observable by which we can learn about the ocean.

Similarly to planetary systems, subsurface ocean measurements in present and past polar regions can also be challenging. In the Arctic, subsurface measurements are significantly easier, particularly with the advent of the Ice-Tethered Profiler, which can measure ocean temperature and salinity below sea ice [24]. In Antarctica, where ice shelves are significantly thicker than Arctic sea ice, observing below remains a challenge. Many recent studies have proposed ways to infer oceanic properties (like melt rate) from ice properties [e.g., 14], on which the planetary literature [23] builds. Further, advanced systems like IceFin [25] are also utilized to observe below ice, and have likewise provided insights on understanding planetary ice systems [e.g., 26]. In past climates where ice sheets or sea ice extended over the present-day ocean, ocean measurements are often inferred from isotopic signatures [27–29], relying on similar methods which also arise in the dating of planetary sample returns [e.g., 30].

## 3 Solution: Link Ice Observations to the Ocean

Thus, we endeavor to understand what properties of the ocean we can infer from ice observations. This applies in planetary, paleo, and present climate settings, where in-situ observables are lacking. Of primary interest is the ocean stratification, which controls the ocean dynamics [31], as well as the mixing of biosignatures and nutrients, which relate to habitability and life [e.g., 32].

So what is the simplest possible inference we can make with an observation that we might actually have, or have imminently? In planetary settings, we can infer oceanic heat fluxes and freeze/melt rates utilizing upcoming ice measurements [e.g., 15, 23], under a number of simplifying assumptions and with a knowledge of ice viscosity (which unfortunately is not well-known). Similar ideas exist across Antarctic science, where basal melt rates are often inferred from conservation of mass and a knowledge of ice thicknesses [e.g., 14, 33]. But why should we care about heat flux or melt rate at

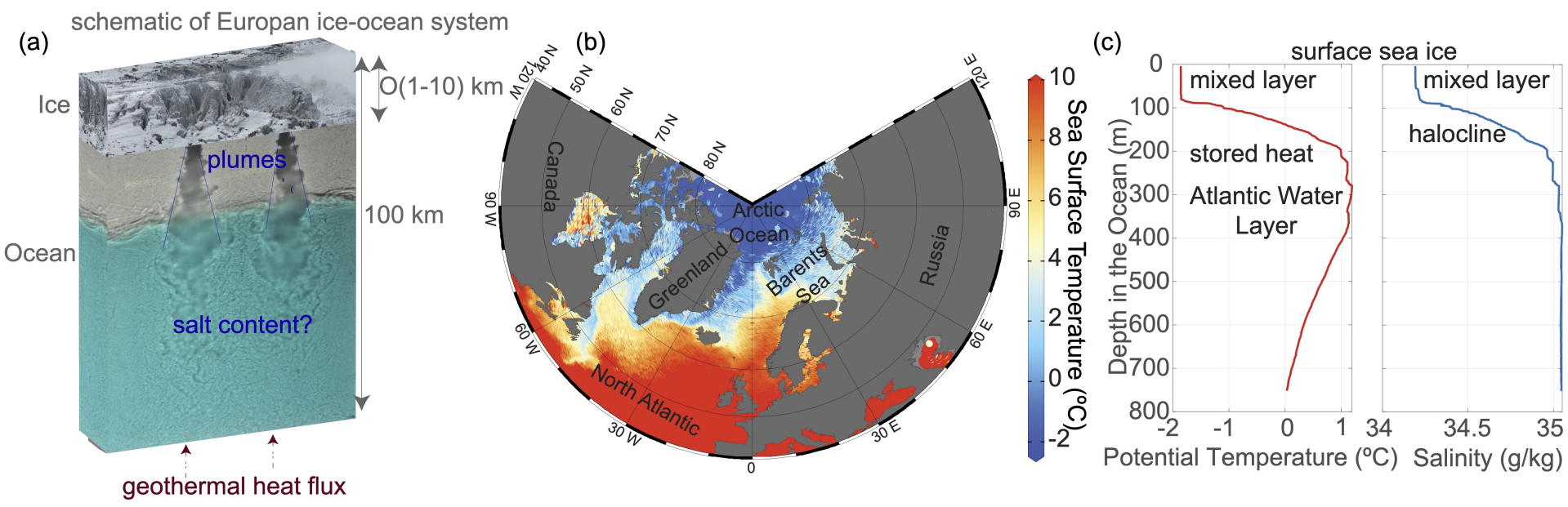

**Fig. 2** (a) Schematic of Europan ice-ocean system [related image in 11]. (b) Map of sea surface temperature (°C) showing the transition between the temperature-stratified North Atlantic Ocean and the salt-stratified Arctic Ocean; (c) Potential temperature (left) and salinity (right) profiles with depth from location near the North Pole. Different water-mass characteristics are labeled.

all? Because they are related to the ocean stratification. Ice melting means imposing freshwater on the ocean, thereby stratifying it. If we know the spatial dependence of where ice melts more, we have some zero-th order way to infer something from an observation (ice thickness) that we actually have, or will have [15].

Of course, ocean mixing also influences the ice cover. If we knew how much ocean mixing was happening and where, from in-situ measurements, we could reasonably make inferences about the ice. For example, in Arctic settings when surface ocean mixing increases, such as from wind-driven mixing, we expect a reduction in ice cover [e.g., 34]. Further, we know that in Antarctica, transport of heat and salt beneath ice sheets influences their topography [e.g., 35, 36], and that during Snowball Earth, for example, oceanic mixing influences the ice cover as well [e.g., 37]. In non-Earth planetary settings, we know even less about subsurface ocean mixing; interpretations of ice shell responses to ocean mixing generally stem from large-scale models [e.g., 38, 39]. Here, often a single mixing diffusivity is prescribed, though depth-dependent or spatially-varying diffusivities may be more realistic. The influence of tidal heating and dissipation also arises in non-Earth planetary settings [e.g., 38], further complicating the results.

# 4 Role of Salinity

In cold and salty-enough oceans, the stratification, and therefore the mixing, is governed by the salt rather than the temperature [e.g., 39, 40]. Therefore, our understanding of ocean salinity when thinking about any of these cold planetary systems is of fundamental importance, since the salt content controls both the circulation and ocean mixing/heat transport (which are related).

Many observational and numerical studies in planetary and polar settings, both Arctic and Antarctic-related, demonstrate this idea. For example, in double-diffusive systems, which manifest in Earth [41–43], planetary [44], and astrophysical environments [45] the heat transport is regulated by the salt stratification [see, e.g., 46, for a theoretical perspective]. Many Solar System studies have addressed the finding that both the magnitude of salinity (which determines whether an ocean is salty enough to be salt-stratified) along with the value of salt stratification is key to icy moon ocean dynamics [e.g., 11, 32, 39, 47, 48] (Figure 2a). In the Arctic, perhaps the idea is most explicitly presented by [49], which introduces a metric to characterize the Arctic halocline's importance to vertical heat fluxes. Recent research on scalings for the influence of salinity and thermal gradients on convective heat fluxes presents relevant applications to the upper layer of Earth's polar oceans as well as other planetary/astrophysical systems [50, 51]. And observations and simulations from the Antarctic region further demonstrate the effects of salt and heat, and their importance to regulating the ice cover [e.g., 35, 36, 52]. Many further examples exist.

## 5 An Integrated Approach

What does this have to do with habitability and life? Salt content is not only relevant for understanding mixing and stratification, and the feedback onto ice cover, in polar or planetary oceans, but also for whether an icy system is biologically-hospitable [53, 54]. For example, in Arctic systems, very salty brine in sea ice gives rise to the conditions for a variety of ecosystems [55]. Since the lateral extent of sea ice is controlled by the salinity below (as sea ice can only grow when an ocean is salt-stratified, as in the Arctic, [40, 56], see Figure 2b,c), this means that a change in oceanic stratification from a thermally-governed to salinity-governed system will also affect biological life on Earth. From the more "climate-dynamics" perspective, ocean stratification places a control on ocean dynamics and mixing [e.g., 57], controlling nutrient or biosignature transport [32, 58]. And via salinity's effect on heat transport, this regulates the thickness and maintenance of an ice cover which keeps our planet cool and other planets warm.

In planetary settings, a more comprehensive understanding of stratification and mixing could inform where to look for life if we are to find it in these systems. In paleo settings, where ocean mixing metrics are not particularly well-coupled with isotopic proxies of mixing [e.g., 57], we wonder whether mixing and the effect on ventilation and circulation may influence our understanding of how complex life arose on Earth. But what if we only have isolated data of salinity or ice thickness? Perhaps AI-based inferences of salinity stratification from sparse in-situ polar, or planetary ice, data could be useful in inferring salinity or ice thickness profiles, building on the framework in [59]. Then, canonical, statistical and machine learning techniques could be synchronously developed to fill in the gaps [as in e.g., 57, 60, for turbulence metrics].

Further, the idea we synthesize here across fields to utilize above-ground ice observations to infer subsurface mixing metrics could be coupled with research on deep ocean mixing [e.g., 57] to complement our knowledge of interior ocean mixing. In cold planetary oceans, we expect that this mixing is largely influenced by the salt. The integration of paleo, polar, and planetary science, and their relationships to ocean mixing and salinity, may help us better answer these important scientific questions as they pertain to habitability and life in the Solar System.

**Acknowledgments.** The MEDEA imagery in Figure 1 came from the USGS Global Fiducials Library. NCS acknowledges funding from the Isaac Newton Joint Schools Research Grants Scheme for the Schools of Technology and Physical Sciences for the proposal "Influence of Salinity on Ice-Ocean Processes." The work largely derives from a presentation by NCS at the 2026 Ocean Mixing Gordon Research Conference.